\documentclass[12pt,preprint]{aastex}

\usepackage{amsmath}
\usepackage{lscape}

\begin{document}

\title{Evolution of Transient Low-Mass X-ray Binaries to Redback Millisecond Pulsars}

\author{Kun Jia$^{1}$ and Xiang-Dong Li$^{1,2}$}

\affil{$^{1}$Department of Astronomy, Nanjing University, Nanjing 210046, China}

\affil{$^{2}$Key laboratory of Modern Astronomy and Astrophysics (Nanjing University), Ministry of
Education, Nanjing 210046, China}

\affil{$^{}$lixd@nju.edu.cn}

\begin{abstract}
Redback millisecond pulsars (hereafter redbacks) are a
sub-population of eclipsing millisecond pulsars in close binaries.
The formation processes of these systems are not clear. The three
pulsars showing transitions between rotation- and accretion-powered
states belong to both redbacks and transient low-mass X-ray binaries
(LMXBs), suggesting a possible evolutionary link between the them.
Through binary evolution calculations, we show that the accretion
disks in almost all LMXBs are subject to the thermal-viscous
instability during certain evolutionary stages, and the parameter
space for the disk instability covers the distribution of known
redbacks in the orbital period - companion mass plane. We
accordingly suggest that the abrupt reduction of the mass accretion
rate during quiescence of transient LMXBs provides a plausible way
to switch on the pulsar activity, leading to the formation of
redbacks, if the neutron star has been spun up to be an energetic
millisecond pulsar. We investigate the evolution of
redbacks, taking into account the evaporation feedback, and discuss
its possible influence on the formation of black widow millisecond pulsars.

\end{abstract}


\section{Introduction}
Millisecond pulsars (MSPs) are a population of neutron stars (NSs)
with fast spins and weak magnetic fields \citep{Backer1982}. They
are thought to have evolved from low-mass X-ray binaries (LMXBs)
\citep[][for a review]{Bhattacharya1991}. The discovery of coherent
millisecond X-ray pulsations from several NSs in LMXBs strongly
confirms this theoretical expectation \citep[][and references
therein]{Patruno2012}. More recent observations demonstrate
transitions between a rotation- and an accretion-powered state in
three systems, i.e.,  PSR J1023$+$0038
\citep{Archibald2009,Stappers2014,Patruno2014}, PSR J1824$-$2452I/IGR
J18245$-$2452 \citep{Papitto2013}, and XSS J12270$-$4859 \citep{de
Martino2010,de Martino2013,Roy2014}.

Black widow and redback MSPs (hereafter redbacks) are two
sub-populations of eclipsing MSPs with orbital periods $P_{\rm orb}
\lesssim 1\rm~day$ \citep[for a recent review]{Roberts2013}. While
redbacks have relatively more massive companions ($0.2\ M_{\odot}
\lesssim M_{2} \lesssim 0.4\ M_{\odot}$), the companion masses in
black widow binaries are significantly lower ($0.02\ M_{\odot}
\lesssim M_{2} \lesssim 0.05\ M_{\odot}$)
\citep{Fruchter1988,Stappers1996}. The most striking feature of both
types of binaries is their regular radio eclipses around superior
conjunction \citep[][and references therein]{Roberts2013}. These deep eclipses indicate a
low-density, highly-ionized gas cloud enwrapping the companions, but
the origin of the eclipsing material is not clear. Consider the
compact orbits and the fact that they are associated with
$\gamma$-ray sources, the eclipsing material could originate from
the companions evaporated by the high-energy particles from the MSPs
\citep{van den Heuvel1988,Ruderman1989a}. Alternatively, the nearly
Roche-lobe (RL) filling companions \citep{van
Kerkwijk2011,Deller2012} suggest that there may be Roche-lobe
overflow (RLOF) in these binaries, and the overflowing matter is
stopped and blown away by the MSP's radiation pressure at the inner
Lagrangian point \citep{Ruderman1989a,Burderi2001,Burderi2002}.

Notably, all the three transitional MSPs discovered so far are
redbacks, which have a currently known population of 17 sources \citep[see Table 1
in][]{Smedley2015}. With a systematic X-ray study of eight nearby
redbacks, \cite{Linares2014b} defined three (pulsar, disk, and
outburst) states according to their X-ray luminosities. In the
pulsar state, the X-ray luminosities are in the range of
$10^{31}{\rm~ergs}< L_{\rm X} < 4\times10^{32} \rm~erg~s^{-1}$, and
they exhibit radio eclipses and pulsations \citep{Archibald2010}. In
the disk state with $4\times10^{32}{\rm~ergs}^{-1} < L_{\rm X} <
10^{34} \rm~erg~s^{-1}$, the disk lines can be detected in the
optical band accompanied with fast bimodal switching in X-rays
\citep{Ferrigno2014,Linares2014a}. The (full accretion) outburst
state with $L_{\rm X} > 10^{34} \rm~erg~s^{-1}$ has been detected
only in PSR J1824$-$2452I so far \citep{Papitto2013}.

In the general picture of the NS$-$accretion disk interaction, the
inner radius of a geometrically thin accretion disk is truncated at the
magnetospheric radius where the energy of the NS magnetic field
equals the kinetic energy of the infalling matter in the disk
\citep{Pringle1972}: $R_{\rm m} \simeq \mu^{4/7}(2GM_{\rm
NS})^{-1/7}\dot{M}^{-2/7}$, where $\mu$ is the magnetic moment of
the NS, $M_{\rm NS}$ the NS mass, $\dot{M}$ the accretion rate, and
$G$ the gravitational constant. With a high enough accretion rate,
the magnetospheric radius can be inside the corotation radius at
which the Keplerian angular velocity in the disk is equal to the angular velocity $\Omega_{\rm NS}$ of the NS:
$R_{\rm co} =(GM_{\rm NS}/\Omega_{\rm NS}^{2})^{1/3}$, and the
accreting matter can be channeled by the field lines to reach the NS
surface. If the accretion rate is reduced and the magnetospheric
radius expands outside the corotation radius, a significant fraction
of the accreting matter may be ejected from the system with the
so-called propeller effect \citep{Illarionov1975}. Furthermore, if
the accretion rate is so low that the magnetospheric radius is
beyond the light cylindrical radius (where the corotating velocity
matches the speed of light, $R_{\rm LC} = c/\Omega_{\rm NS}$), the
disk is disrupted and a radio pulsar switches on. So, to account for
the mass outflows and transitional behavior in redbacks, we require
a significant variation in the mass accretion rate during the binary
evolution.

Based on this argument, three models have been proposed for the
formation of redbacks by invoking interrupted mass transfer
processes. Below we briefly review these models.

{\noindent\em 1. The disrupted magnetic braking (MB) model}

Mass transfer in short orbital period LMXBs is driven by angular
momentum loss (AML) due to MB and gravitational radiation (GR). When
its mass decreases to around $0.2-0.3\ M_{\odot}$, the donor becomes
fully convective and the MB-induced AML is greatly reduced, leading
to RL decoupling \citep[e.g.][]{Spruit1983}. As a result, the mass
transfer is temporarily halted. \cite{Chen2013} assumed that a MSP
then switches on and begins to evaporate its companion star with high-energy radiation. They investigated the subsequent binary evolution, and showed that adopting different values of the evaporation efficiency  could account for the formation of both
black widows and redbacks. In this model it is hard to explain a few
redbacks with relatively large companion masses ($\sim 0.5\
M_{\odot}$) and long orbital periods ($\gtrsim 0.5$ day). The
authors suggested that these systems may have been temporarily
detached earlier for some reasons, then turned into the evaporation
stage with a relatively massive companion star.

{\noindent\em 2. The irradiation-induced cyclic mass transfer model}

When the donor star in a LMXB is irradiated by the X-rays from the
accreting NS, its intrinsic luminosity can be blocked by irradiation
\citep{Podsiadlowski1991,Hameury1993}. Due to the thermal relaxation
of the convective envelope of the donor star, the secular mass
transfer could be unstable under some conditions, cycling between
low and high states \citep{Ritter1996,King1995,King1996,Buning2004}.
\cite{Benvenuto2014,Benvenuto2015} found that this irradiation
instability and cyclic mass transfer processes are popular during
the evolution of contracting LMXBs, and they suggested that redbacks
may be formed during the low state of the mass transfer. In this
scenario, the redback companions should almost fill their RL. By
considering the evaporation due to the MSP's wind/radiation, they
found that black widows can descend from redbacks, but not all
redbacks evolve into black widows. A caveat is that this model may
not work in transient LMXBs, in which the mass transfer cycles would
be suppressed due to intermittent irradiation \citep{Ritter2008}.

{\noindent\em 3. The accretion-induced collapse (AIC) model}

\cite{Smedley2015} suggested that redbacks may be formed during the
AIC of a white dwarf (WD) in cataclysmic variable (CV)-like systems
besides the traditional recycling scenarios. At the time of AIC, the
companion star becomes detached from its RL because of the
orbital expansion caused by sudden gravitational mass loss. The mass
transfer halts and the newborn MSP starts to ablate its companion.
Like the cyclic mass transfer model, the AIC model can reproduce all
redbacks in the $M_{2}-P_{\rm orb}$ plane. However, \cite{Ablimit2015}
pointed out that it is difficult or impossible to produce the requisite occurrence of
AIC in traditional CVs, unless irradiation-excited wind from the donor star is also taken into account.

More recently, channeled accretion was detected in PSR J1023$+$0038
\citep{Archibald2015}, which is actually an accreting millisecond
X-ray pulsar (AMXP). Unlike other AMXPs, this channeled accretion
was discovered at a very low X-ray luminosity $L_{X}~<~10^{34}~\rm
erg~s^{-1}$, which challenges the traditional accretion-propeller
theory \citep{Archibald2015,Bogdanov2014}. Similar phenomenon  has
also been discovered in XSS~J12270$-$4859 \citep{Papitto2015}.
Adding PSR J1824$-$2452I as a typical AMXP in the outburst state
\citep{Papitto2013}, we note that all the three transitional
redbacks exhibit as AMXPs at the disk or outburst state. As we know,
all the AMXPs are transient systems \citep{Patruno2012}, likely
caused by the thermal-viscous instability in the accretion disks
\citep[see][for a review]{Lasota2001}. It is also noted that,  all
NS-LMXBs are subject to the disk instability either with evolved
companions \citep{King1997} or with low main-sequence (MS)
companions after the cessation of MB, as we show below. Observationally
nearly half of the LMXBs in the Galactic disk are transient sources
\citep{Ritter2003,Knevitt2014}. Based on these features, we consider
the thermal-viscous disk instability as  a possible mechanism for the formation of
redbacks, and investigate their evolutionary sequences.

The rest of this paper is organized as follows. In Section 2  we
briefly introduce the binary evolution model. Our main results on
the parameter space of the disk instability and its effect on the
LMXB evolution are presented in Sections 3 and 4, respectively. They
are then discussed in Section 5 and summarized in Section 6.

\section{Method}

We calculate the LMXB evolution with an updated version
(7624) of the Modules for Experiments in Stellar Astrophysics (MESA)
\citep{Paxton2011,Paxton2013,Paxton2015}. The binary is
initially composed of a NS (of mass $M_{1}$) and a zero-age main-sequence
(ZAMS) companion star (of mass $M_{2}$) with solar chemical
compositions.
We use the \cite{Eggleton1983} formula to calculate the effective RL radius of the companion (or donor) star,
\begin{equation}
\frac{R_\mathrm{L,2}}{a}=\dfrac{0.49q^{-2/3}}{0.6q^{-2/3}+\ln(1+q^{-1/3})},
\end{equation}
where $q=M_{1}/M_{2}$ is the mass ratio and $a$ is the orbital
separation. To calculate the mass transfer rate via RLOF we
adopt the Ritter scheme \citep{Ritter1988,Paxton2015} in the code,
which takes into account the finite scale height of the stellar atmosphere of the donor,
\begin{equation}
-\dot{M}_2=\dot{M}_0\exp\left[\frac{R_2-R_{\rm L, 2}}{H_{\rm
P}/\gamma(q)}\right].
\end{equation}
Here $\dot{M}_0$ depends on the mass ratio, the density at the
donor's photosphere and its effective temperature, $R_2$ is the
donor's radius, $H_{\rm P}$ is the pressure scale height of the atmosphere, and $\gamma$ is a function of $q$.
The rate of AML during the
evolution consists of three parts:
\begin{equation}
\dot{J}=\dot{J}_\mathrm{GR}+\dot{J}_\mathrm{ML}+\dot{J}_\mathrm{MB}.
\end{equation}
The first term $\dot{J}_\mathrm{GR}$ on the right hand side is the
rate of GR-induced AML given by \citep{Landau1975},
\begin{equation}
\dot{J}_\mathrm{GR}=-\dfrac{32}{5}\dfrac{G^{7/2}}{c^{5}}\dfrac{M_{1}^{2}M_{2}^{2}(M_{1}+M_{2}^
{1/2})}{a^{7/2}}.
\end{equation}
with $c$ the speed of light.
The second term is due to the mass loss from the system.  We assume
that a fraction ($\beta$) of the transferred mass is accreted by
the NS and the rest is ejected out of the system by isotropic winds,
taking away the specific AM of the NS,
\begin{equation}
\dot{J}_\mathrm{ML}=-\left(1-\beta\right)\dot{M}_{2}\left(\dfrac{{M}_{2}}{{M}_{1}+{M}_{2}}\right)^{2}a^{2}\omega,
\end{equation}
where $\omega$ is the angular velocity of the binary orbit. We also
assume that the accretion rate $\dot{M}_1$ is limited by the
Eddington critical rate,
\begin{equation}
\dot{M}_{\rm Edd}=3.6\times10^{-8}\left(\frac{M_{1}}{1.4M_{\odot}}\right)\left(\frac{0.1}{GM_{1}/R_{1}c^{2}}\right)\left(\frac
{1.7}{1+X}\right)M_{\odot}\rm yr^{-1},
\end{equation}
where $X$ is the hydrogen abundance and $R_{1}$ is the NS radius, taken to be $10^{6}~\rm cm$.  Our
previous work shows that the value of $\beta$ does not
significantly affect the evolutionary tracks of LMXBs
\citep{Jia2014}, so here we fix its value to be 0.5. Thus,
$\dot{M}_1=\min(-\dot{M}_2/2,\dot{M}_{\rm Edd})$.

The last term in Eq.~(3) represents the AML due to MB, for which we adopt the
standard formula \citep{Verbunt1981,Rappaport1983},
\begin{equation}
\dot{J}_\mathrm{MB}=-3.8\times10^{-30}M_{2}R_{2}^{4}\omega^{3}\ {\rm
dyn\,cm}.
\end{equation}
We assume that MB is reduced when the star's convective envelope is
very thin, and add an ad hoc factor to $\dot{J}_\mathrm{MB}$
\citep{Podsiadlowski2002},
\begin{center}
$\exp(-0.02/q_\mathrm{conv}+1)$\ \ (if $q_\mathrm{conv}<0.02)$,
\end{center}
where $q_\mathrm{conv}$ is the mass fraction of the surface convective envelope. We also terminate MB when the central stellar radiative zone vanishes.

The accretion disk in a LMXB becomes thermally and viscously
unstable when the mass transfer rate drops below a critical value.
Then the LMXB may appear as a transient source experiencing limit
cycles between short outbursts and long quiescent intervals with a
duty cycle $d\sim0.001-0.1$ \citep{King2003}. Here we adopt the
critical mass transfer rate for an irradiated disk given by
\citep{Dubus1999},
\begin{equation}
\dot{M}_{\rm cr}\simeq3.2\times10^{-9}\left( \dfrac{M_{1}}{1.4M_{\odot}}\right)^{0.5}\left(\dfrac{M_{2}}{1.0M_
{\odot}}\right)^{-0.2}\left(\dfrac{P_{\rm orb}}{1.0\,\rm d}\right)^{1.4}M_{\odot}\rm yr^{-1}.
\end{equation}
Once $\dot{M}_1<\dot{M}_{\rm cr}$, we simply assume  that there is
no accretion during quiescence, and the matter flow accretes at a
rate of $\dot{M}_1=\min(-\dot{M}_{2}/2d,\dot{M}_{\rm Edd})$  during
outbursts. Here we set the duty cycle $d=0.01$ throughout this
paper. Since the evolution time steps in the code are
usually much longer than the disk outburst cycle time, the average
mass transfer rate from the donor star actually remains the same as
in the case of stable disk accretion, but the accreted mass by the
NS is considerably reduced.

\section{The parameter space for unstable disks}
We calculate the evolution of a grid of LMXBs with a
$1.35\,M_{\odot}$ NS and a donor star of initial mass $M_{2, \rm i} = 1.0, 1.5, $
and $2.0~M_{\odot}$. The initial orbital periods $P_{\rm orb, i}$ is set to range
from $\sim 0.4$ day to $\sim 20$ days, to cover the evolution of CV-like, ultra-compact and wide LMXBs \citep{Deloye2008}.

We show some of the calculated evolutionary tracks in the
$M_{2}-P_{\rm orb}$ plane (Fig.~1). From left to right, the three
panels correspond to $M_{2, \rm i} = 1.0, 1.5, $ and $2.0~M_{\odot}$,
respectively. In each panel the green solid and red dotted lines
represent the mass transfer processes with stable and unstable
accretion disks, respectively. The black dashed lines denote the stage of
RL detachment (i.e., without mass transfer). The asterisk on each
evolutionary track indicates that the accreted mass reaches
$0.1\,M_{\odot}$, a typical amount to spin up a NS to
millisecond periods \citep{Lipunov1984,Zhang2011,Taurisetal2012}.
The dot-dashed line shows the relation between the orbital period
and the (WD) core mass of the donor (i.e., $P_{\rm orb}-M_{\rm WD}$
relation) after Case B mass transfer
\citep{Joss1987,Rappaport1995,Tauris1999,Lin2011,Jia2014}. Here we
use the calculated results of \citet{Lin2011}. The two brown dashed lines
give the orbital periods when the donor star is on ZAMS and
terminal-age main-sequence (TAMS). They are obtained by combining
the orbital period - donor radius  relation \citep{Rappaport1995}
\begin{equation}
P_{\rm orb} = 20 G^{-1/2} R_{2}^{3/2} M_{2}^{-1/2},
\end{equation}
with the mass - radius relations  \citep{Demircan1991} for ZAMS stars,
\begin{equation}
R_{2}\simeq\left\{
\begin{aligned}[l]
0.89(M_{2}/M_{\odot})^{0.89}~R_{\odot}\ \ ({\rm for}\ M_2<1.66M_{\odot})\\
1.01(M_{2}/M_{\odot})^{0.57}~R_{\odot}\ \ ({\rm for}\ M_2>1.66M_{\odot})\\
\end{aligned}
\right.,
\end{equation}
and for TAMS stars,
\begin{equation}
R_{2}\simeq\left\{
\begin{aligned}
2.00(M_{2}/M_{\odot})^{0.75}~R_{\odot}\ \ ({\rm for}\ M_2<1.66M_{\odot})\\
1.61(M_{2}/M_{\odot})^{0.83}~R_{\odot}\ \ ({\rm for}\ M_2>1.66M_{\odot})\\
\end{aligned}
\right.,
\end{equation}
respectively.

In Fig.~1 we also plot the known redbacks with open diamonds and
suqares in the Galactic field (field redbacks) and globular
clusters (GC redbacks) listed in \cite{Smedley2015}, respectively.
Note that we have recalculated their masses by assuming a NS mass of
$1.35~M_{\odot}$ instead of $1.25~M_{\odot}$ in \cite{Smedley2015}.

\cite{Iben1995} divided the LMXB evolution into 7 cases according to
the evolutionary state of the donor star and the mass transfer mode.
Our calculations demonstrate that the LMXBs are persistent, i.e., non-transient, in the following cases when the mass transfer is driven by (1) MB for CV-like LMXBs with MS
donors; (2) thermal expansion of a massive evolved donor; and (3) GR in
ultra-compact systems with $P_{\rm~orb} \lesssim1\rm~hour$.

Figure~2 shows some illustrative examples of mass transfer rates in the black solid lines. We divide the evolutionary tracks into three
types according to whether the initial orbital period being longer or shorter than, or close to the so-called the "bifurcation period"
\citep{Pylyser1988,Pylyser1989}. The red dashed lines depict the
critical mass transfer rates for comparison.

The bottom row of Fig.~2 shows the evolutions of
contracting systems. In the case of $M_{2, \rm
i}=1.0~M_{\odot}$ and $P_{\rm orb,i}=0.4$ day, the mass transfer
begins when the companion star is still unevolved, so MB dominates
the AML and the mass transfer rate is initially above $\dot{M}_{\rm cr}$. When
the donor becomes fully convective, MB stops and AML is solely
caused by GR, the mass transfer rate drops down below $ \dot{M}_{\rm
cr}$ and the binary becomes transient. When $M_{2, \rm i}=1.5~M_{\odot}$
and $P_{\rm orb,i}=0.5$ day, in the initial stage of the mass
transfer, MB takes no effect because the donor has a radiative
envelope, and the accretion disk is unstable because of the low mass
transfer rate. When the stripped donor star has developed a
convective envelope, MB starts to work and the binary becomes
persistent. The following evolution is similar to the case of
$M_{2}=1.0~M_{\odot}$. For a more massive donor with
$M_{2, \rm i}=2.0~M_{\odot}$ and $P_{\rm orb,i}=0.55$ day, mass transfer proceeds rapidly on a thermal timescale \citep
{Podsiadlowski2002,Lin2011}. After that the mass transfer rate
decays. When the donor mass is comparable to the NS mass before MB
takes effect, the mass transfer rate drops below $ \dot{M}_{\rm cr}$
and produces a dip. The subsequent evolution is also similar to the
previous cases.

The evolutions of widening binaries are shown in the top row of
Fig.~2. When $M_{2, \rm i}=1.0~M_{\odot}$ and $P_{\rm orb,i}=20.0$ days,
different from the contracting binaries, the donor star is evolved and the mass transfer is driven
by the stellar expansion due to shell burning on the red giant
branch, which results in a expanding orbit and a much higher
$\dot{M}_{\rm cr}$. Accordingly, we see that the mass transfer rate
is always below $\dot{M}_{\rm cr}$. In the cases of $M_{2, \rm i}
=1.5~M_{\odot}$, $P_{\rm orb,i}=10.0$ days, and $M_{2, \rm i}=2.0~M_{\odot}$,
$P_{\rm orb,i}=2.0$ days, a mass transfer rate above
$10^{-7}M_{\odot}\rm yr^{-1}$ is reached at the initial stage. As
the evolution proceeds, the mass transfer rate descends while
$\dot{M}_{\rm cr}$ increases due to orbital expansion, so the disks
become unstable for wider systems. The final evolutionary products are MSP-He
WD binaries.

The middle row of Fig.~2 presents examples just between
contracting and widening binaries, with the final
orbital periods around 1 day. At the beginning of mass transfer, the
donor star is on the MS, similar as in contracting binaries,
and the accretion disk is stable. Then the mass transfer rate
gradually declines to be below $\dot{M}_{\rm cr}$, due to the
slight expansion of the donor during the MS evolution. For
$M_{2, \rm i}=2.0~M_{\odot}$, there is a spike in the mass transfer
rate at $M_{2}\sim0.5-0.6~M_{\odot}$, which indicates the end of
the MS evolution. The donor star evolves rapidly to pass through the
Hertzsprung gap, when its H-exhausted core mass exceeds the
Sch\"{o}nberg-Chandrasekhar limit. In the case of $M_{2, \rm
i}=1.0~M_{\odot}$, there is not such a spike since the
Sch\"{o}nberg-Chandrasekhar limit is irrelevant for lower mass
($<1.1M_{\odot}$) stars, in which electron degeneracy pressure
dominates the He core. These  stars can gradually transit from
central- to shell-hydrogen burning with no Hertzsprung gap in the
H-R diagram.

Go back to Fig.~1,  we see that the parameter space for the
occurrence of the disk instability covers the distribution of known
redbacks. If MSP activity switches on during the quiescent phase of LMXBs
(as observed in the three transitional pulsars), it may lead to the formation of redbacks.

\section{Evolution with evaporation}

Considering the fact that the evaporation processes can be rather complicated, here we
adopt the simple picture that, when a NS in a LMXB spins at
millisecond periods and the mass transfer rate declines to be low
enough, a radio pulsar will turn on and prevent further accretion
\citep{Ruderman1989a,Burderi2001,Burderi2002}; meanwhile, part of
the spin-down energy is used to evaporate the companion star
\citep{van den Heuvel1988}. Therefore, we assume that evaporation in
LMXBs takes place once the following two conditions are satisfied:
(1) during the mass transfer phase, a NS has accreted at least
$0.1M_{\odot}$ mass to be recycled as a MSP, and (2) the mass
accretion rate decreases significantly, either due to a decrease in
the mass transfer rate or because the accretion disk becomes
unstable.

We note that, if the disk is in an unstable state,
evaporation can take place effectively simultaneously as far as the
evolution time steps in the code are concerned. Our calculations
show that there can be either RLOF or not along with evaporation,
depending on the subsequent evolution of the stellar radius and the
RL radius (see discussion in next section). In the case of evaporation accompanied with RLOF, we assume that the accretion disk is
outside the light cylinder (at $\sim100$ km) during the quiescent phase, then the NS
does not accrete and acts as a radio pulsar. The material in the
inner disk is blown away from the binary, but most of the
transferred material from the companion star continues to accumulate
in the outer region of the disk ($\sim10^{2}-10^{6}$ km). When the disk switches to the
outburst phase, the disk material rapidly accretes onto the NS and the pulsar shuts off.
Since the accretion rates during outbursts are generally
super-Eddington, most of the material is also assumed to be lost from the binary.
Thus, considering the long evolution time steps in the calculation,
we assume that almost all of the transferred matter is ejected,
taking away the specific AM of the NS.

In the mean time, to explore the evaporation effect on the secular
evolution, we assume that evaporation proceeds at a rate \citep{van
den Heuvel1988,Stevens1992},
\begin{equation}
\dot{M}_{2,\rm evap}=-\dfrac{f}{2v_{2,\rm esc}^{2}}L_{\rm
p}\left(\dfrac{R_{2}}{a}\right)^{2},
\end{equation}
where $v_{2,\rm esc}$ is the escape velocity at the surface of the
companion star, $L_{\rm p}=4\pi^ {2}I\dot{P}/P^{3}$ is the NS's
spin-down luminosity, and $f$ is the efficiency of evaporation. Here
$I$, $P$, and $\dot{P}$ are the moment of inertia, the spin period
and its derivative of the NS, respectively. We assume that the
evaporating material takes away the specific AM of the companion
star.

When a radio pulsar turns on, we assume that its spin evolves
following the standard magnetic dipole radiation model with a
constant braking index $n = 3$ \citep{Shapiro1983}, and neglect the
change in $P$ due to accretion during outbursts because of their
short duration. We take typical values of the initial spin period
$P=3$ ms, period derivative $\dot{P}=1.0\times10^{-20}\,\rm
ss^{-1}$, and moment of inertia $I=10^{45}\,\rm gcm^{2}$.

Similar to Fig.~1, we show a series of evolutionary sequences in
Fig.~3 taking into account the evaporation effect, where we set the
efficiency factor $f = 0.1$.  The beginning of evaporation
is denoted with a pentagram on each evolutionary track. In some
cases it starts due to a drop of the mass transfer rate when the
companion mass is comparable with the NS mass, as shown in Fig.~4.
Comparison between Figs.~1 and 3 demonstrates that evaporation can
considerably alter the evolution of LMXBs, in broad agreement of
\cite{Chen2013} and \cite {Smedley2015}. However, the orbital period
evolution in Fig.~3 is more flattened than in \cite{Chen2013} and
\cite {Smedley2015}, since we take account of the AML due to the
evaporating material. This feature can be illustrated as follows.
The orbital angular momentum of the binary is
\begin{equation}
J=M_1M_2\left(\frac{Ga}{M}\right)^{1/2},
\end{equation}
where $M=M_1+M_2$ is the total mass. Logarithmically
differentiating the above equation with respect to time gives the
change in the orbital separation due only to evaporative mass loss
from the companion star,
\begin{equation}
\frac{\dot{a}}{a}=-\frac{2\dot{M}_2}{M_2}\left(1-\frac{M_2}{2M}-\alpha_{\rm evap}\frac{M}{M_1}\right),
\end{equation}
where $\alpha_{\rm evap}$ is the specific AM of the evaporative wind in units of the binary's specific AM. If  the evaporative wind carries the
specific AM of the companion star, we have
\begin{equation}
\frac{\dot{a}}{a}=-\frac{2\dot{M}_2}{M_2}\left(1-\frac{M_2}{2M}-\frac{M_1}{M}\right).
\end{equation}
In the limit where
$M_2\ll M_1$, Eq.~(14) becomes
\begin{equation}
\frac{\dot{a}}{a}\simeq -\frac{2\dot{M}_2}{M_2}(1-\alpha_{\rm evap}).
\end{equation}
Thus if $\alpha_{\rm evap}\rightarrow 0$ (i.e., no AML due to
evaporative mass loss) the orbit will expand considerably, while if
$\alpha_{\rm evap}\rightarrow 1$ the orbit will not expand at all.

In Fig.~4 we display how evaporation influences the binary evolution
for the three cases in Fig.~2. The black solid and blue dotted lines
represent the RLOF mass transfer rate and the evaporative wind loss
rate, respectively. We start with the CV-like contracting
systems as shown in the bottom row of Fig.~4. Here evaporation takes
place when the donor star decouples from its RL due to the cessation of MB, same as in \cite{Chen2013}, but the
orbital periods of the resultant redbacks are around 0.1 day, different from  in the range of $\sim 0.1-1$ day in \cite{Chen2013}. As mentioned before, for donor stars with
mass $M_{2, \rm i}=1.5$ and $2.0~M_{\odot}$, there exists a dip in
the mass transfer rate (which is below $\dot {M}_{\rm cr}$) before
MB operates. Evaporation may also take place at this time if the NS
has accreted enough mass to be recycled as a MSP. This
suggests the possibility of detecting redbacks with $M_{2}\gtrsim
1.0~M_{\odot}$. When the donor's mass decreases to $\sim1.0~M_{\odot}$, its convective envelope is developed and MB starts to operate, then the mass transfer
increases and evaporation stops. It will start again when the
companion star is detached from its RL.

The top row of Fig.~4 presents the evolutions in the case of
widening systems. Mass accretion of NSs is generally
inefficient in these systems, because of both short evolutionary
time and transient behavior, so the NSs may not be fully
recycled. Nevertheless, for relatively massive donor stars such as
$M_{2, \rm i}=2.0\,M_{\odot}$, the NS can still accrete enough mass
during the intermediate-mass X-ray binary phase \citep{Shao2012},
and we may expect redbacks with subgiant companions and orbital
periods up to tens of days. In the case of
$M_{2,\rm i}=2.0~M_{\odot}$ and $P_{\rm orb,i}=2.0$ days, at the end of
evolution, the binary undergoes a short-lived ($\sim 10^3$ yrs)
episode of additional RLOF as a result of the hydrogen shell flashes
on the proto-WD \citep[e.g.,][]{Antoniadis2014}. The resultant
MSP-He WD binaries follow the $P_{\rm orb}-M_{\rm WD}$ relation
because the companions are generally RL filling. However, all
redbacks discovered so far are in relatively compact orbits with
period less than $\sim 1.5$ day, so there seems to be a gap between
observations and theoretical expectation, and we will return to this
point later.

In the middle row of Fig.~4 the orbital periods of redbacks are
distributed around $0.1-1$ day, close to the bifurcation period.
Different from the case of the contracting systems, the mass transfer
rate drops below $\dot{M}_{\rm cr}$ before MB ceases. Along with
evaporation, the donor star keeps overflowing its RL. The mass
transfer rate declines but the total mass loss rate of the donor
star increases. When the donor star approaches the end of nuclear
evolution, its cannot adjust immediately to fill its RL, RLOF
terminates and evaporation starts to dominate the mass loss. In this case the redback companions may either fill or
underfill their RLs, and the finally formed MSP-He WD binaries may
slightly deviate from the $P_{\rm orb}-M_{\rm WD}$ relation.

In summary, evaporation can considerably influence the secular
binary evolution, especially for relatively compact systems. The
switch-on of radio activity and start of evaporation are caused by
the transient behavior of LMXBs, as well as disrupted MB. The
predicted parameter space is consistent with the distribution of
known redbacks in the $M_{2}-P_{\rm orb}$ plane.

\section{Discussion}

\noindent (1) Transitional pulsars and redbacks

We have shown that transient LMXBs that harbor energetic MSPs may
lead to the formation of redbacks. The three transitional pulsars
clearly demonstrate the close relation between redbacks and
transient AMXPs. Moreover, \cite {Papitto2015} found a very similar
spin distribution between AMXPs and eclipsing MSPs, suggesting that
the two groups belong to a common class at the same evolutionary
epoch. This is also consistent with our hypothesis that both
redbacks and AMXPs originate from transient LMXBs. The related
questions are: which redbacks can experience transitions to the disk
accretion state, and why don't all transient AMXPs show radio
pulsations during quiescence?

Based on the statistics of the current sample, \cite{Linares2014a}
divided redbacks into two groups according to their luminosities
during the pulsar state, and suggested that the bright ones with
X-ray luminosity above $10^ {32}\rm~ergs^{-1}$ are promising
candidates to develop accretion disks.  Considering the fact that disk accretion always requires
RLOF, obviously redbacks with a lobe-filling companion are likely to
experience transitions to the accretion state, while those with a
underfilling companion should stay in the radio pulsar state.
Our results show that the
redback companions can either fill or underfill their RLs, depending on how much mass and AM are lost with evaporation. The condition can be briefly analyzed as follows.
For stable mass transfer via RLOF, from the condition
\begin{equation}
\frac{\dot{R}_2}{R_2}=\frac{\dot{R}_{\rm L, 2}}{R_{\rm L, 2}},
\end{equation}
one can derive the following formula for the mass transfer rate \citep{Rappaport1983}
\begin{equation}
-\dfrac{\dot{M}_{2}}{M_{2}}=\frac{\dfrac{1}{2}\left(\dfrac{\dot{R_{2}}}{R_{2}}\right)_{\rm evol,therm}
-\left(\dfrac{\dot{J}}{J}\right)_{\rm MB,GR}}{\dfrac{5}{6}+\dfrac{\xi_{\rm ad}}{2}-\dfrac{1-\beta}{3(1+q)}-\dfrac{(1-\beta)\alpha(1+q)+\beta}{q}},
\end{equation}
where $(\dot{R_{2}}/R_{2})_{\rm evol,therm}$ represents the change in the companion star's radius under thermal or nuclear evolution,
$\dot{J}_{\rm MB, GR}$ is the AML rate due to MB and GR, $\xi_{\rm ad}$ is the adiabatic radius-mass exponent, $\beta$ is the fraction of the mass overflowing from the donor star that is retained by the NS, and $\alpha$ is the specific AM of the lost material in units of the binary's specific AM. If we
consider RLOF and evaporation simultaneously, we can divide the total mass loss rate of the donor into two parts, $\dot{M_{2}} = \dot{M}_{2,\rm RL} + \dot{M}_{2,\rm evap}$, where $\dot{M}_{2,\rm RL}$ and $\dot{M}_{2,\rm evap}$ are the mass loss rates through RLOF and evaporative wind, respectively. Assuming that the mass transfer is partially non-conservative and the winds from the NS and the donor take away the specific AM of $\alpha_{\rm RL}$ and $\alpha_{\rm evap}$, respectively, we can rewrite Eq.~(18) to be,
\begin{eqnarray}
\dfrac{\vert\dot{M}_{2,\rm RL}\vert}{M_{2}}=
\dfrac{N}{D},
\end{eqnarray}
where
\begin{eqnarray}
N=
\left(\dfrac{\dot{R_{2}}}{2R_{2}}\right)_{\rm evol,therm}
+\left(\dfrac{\vert\dot{J}\vert}{J}\right)_{\rm MB,GR}
-\dfrac{\vert\dot{M}_{2,\rm evap}\vert}{M_{2}}F(q,\xi_{\rm ad},\alpha_{\rm evap}),
\end{eqnarray}
with
\begin{equation}
F(q,\xi_{\rm ad},\alpha_{\rm evap})=\dfrac{5}{6}+\dfrac{\xi_{\rm ad}}{2}
-\alpha_{\rm evap}\dfrac{(1+q)}{q}-\dfrac{1}{3(1+q)},
\end{equation}
and the denominator is,
\begin{eqnarray}
D=
\dfrac{5}{6}+\dfrac{\xi_{\rm ad}}{2}-\dfrac{1-\beta}{3(1+q)}-
\dfrac{\alpha_{\rm RL}(1-\beta)(1+q)+\beta}{q}.
\end{eqnarray}

As pointed out by \cite{Rappaport1983}, stable mass transfer demands both the numerator and denominator of the right hand of Eq.~(20) to be positive. If the denominator $D<0$ then the mass transfer is dynamically unstable, while the system will become detached if the numerator is negative. The evaporative wind can either contribute to, or detract from, the star filling its RL, depending on the AM that it takes away. For stable mass transfer with $D > 0$, the evaporative wind adds a negative term $-\dfrac{\vert\dot{M}_{2,\rm evap}\vert}{M_{2}}F(q,\xi_{\rm ad},\alpha_{\rm evap})$ to the numerator, which tends to make the donor star underfill its RL eventually (as in the case of black widows). However, the details of evaporation are still not clear, in the following we will show how different values of $\alpha_{\rm evap}$ can influence the binary evolution.

Finally, although the three transitional pulsars are only a
small part of AMXPs, it is possible that active radio pulsars have
indeed turned on during the quiescent state of LMXBs, even with no
detected radio pulsations \citep{Iacolina2010}.

\noindent (2) The redback companions

Optical/infrared observations can reveal the evolutionary state of
the redback companions  and the nature of interacting binary
systems. Statistically, most redbacks possess MS companions which is
under strong irradiation with spectral type changes or optical
variability between inferior and superior conjunctions \citep
{Breton2013,Martino2014,Li2014}. Other types of companion stars are
also possible. For example, for the redback PSR J1740$-$5340 in NGC
6397, two scenarios were proposed to explain the inferred large mass
loss rate from its companion, i.e. a MS star ablated by the MSP's impinging flux or inhibited RLOF of an evolved star by the pulsar's
wind flux \citep{D'mico2001a,D'mico2001b,Ferraro2011}.
\cite{Ergma2003} and \cite{Mucciarelli2013} investigated the surface
chemical abundances of its companion, and suggested the pulsar's
companion to be a deeply peeled evolved star. Additionally, PSR
J1816+4510 may possess a (proto-)He WD companion, which is well
within its RL \citep{Kaplan2013}. Therefore, the redback companions
could be MS stars, evolved stars or (proto-)WDs. These features are
compatible with the expectations in our formation scenario of
redbacks.

The condition for the disk instability favors wide systems, and our
calculations suggest that transient MSP-LMXB systems can form with
orbital periods up to tens of days. It is interesting to note that
the longest orbital period of known redbacks is 32.5 hours for PSR
J1740$-$5340. By the way, all discovered AMXPs are distributed in a
similar range of orbital periods, which are less than 1 day
\citep{Patruno2012}. So one would ask why there are no redbacks in
wider orbits. This question is also related to the fact that transient LMXBs with orbital periods longer than 1 day  have thus far been rare \citep{Ritter2003,Knevitt2014}. Probably redbacks
do possess giant companions, but their number is very small due to
their short evolutionary times \citep{Shao2015}.

\noindent (3) Efficiency of irradiation and evaporation

Unlike CVs, there are few LMXBs with orbital period less than 4
hours, since X-ray irradiation of the donor star alters the
long-term evolution of LMXBs \citep{Kolb1996}. It might also be also
responsible for the birthrates discrepancy between Galactic LMXBs
and low-mass binary radio pulsars \citep{Kulkarni1988}.
X-ray irradiation can affect the structure of the donor star by
depressing its intrinsic luminosity. \cite{Podsiadlowski1991} showed
that spherically symmetric irradiation could expand the donor star
and elevate the mass transfer rate, hence shortening the timescale
of mass transfer \citep[see however][]{Nelson2003}.
\cite{Hameury1993} investigated the situation of asymmetrical
irradiation, and found that cyclic mass transfer was possible for a
significant fraction of LMXBs, due to the irradiation instability
\citep[see also][]{Ritter2000,Buning2004}. Recently,
\cite{Benvenuto2014,Benvenuto2015} proposed that the redback
companions should be quasi-RL filling stars during the detached
episode of the cyclic mass transfer.
The occurrence of the irradiation instability requires sustained strong irradiation over at least a thermal timescale of the convective envelope of the companion star, which is typically millions of years for a low-mass MS star and much longer than the thermal-viscous timescale (less than years for compact binaries). The fact that redback systems are distributed in the thermal-viscous instability region, implies the irradiation instability may be suppressed in these
binaries due to the intermittent irradiation \citep{Ritter2008}.

The formation of eclipsing pulsars like black widows and redbacks
also critically depends on the efficiency of evaporation of the
companion star and related AML. In the literature, the proposed energy source of
evaporation can be either the spin-down power of a MSP
\citep{Ruderman1989a} or the accretion power during the LMXB phase
\citep {Ruderman1989b}, and we have only considered the former case
in this work. \cite{van den Heuvel1988} first realized that MSP's
evaporation could account for the lack of LMXBs below the period
gap.
Therefore, to investigate the influence of the evaporation, we perform calculations of the LMXB evolution with $M_{2, \rm i}
=1.0~M_{\odot}$ and different values of $f$\footnote{The change in $f$ reflects both the geometric effect and the
variation in the initial spin-down power, which is fixed in the
calculation.} and AML modes. The results are shown in Fig.~5, in which the line
styles and the symbols have the same meanings as in Fig.~3, and we add
the black widow systems with data taken from A. Patruno's catalogue\footnote{https://apatruno.wordpress.com/about/millisecond-pulsar-catalogue/}. Also note that for ultra-compact systems the critical mass transfer rate for unstable disks should be replaced with those for  He- or C/O-dominated disks \citep{Menou2002,Lasota2008}, but this actually causes very small change of the stable/unstable disk area.

In the upper, middle, and lower panels of Fig.~5, we assume that the evaporative wind takes the specific AM of the donor star (Mode A), at the inner Lagrangian point (Mode B), and in the extreme (non-physical) case with no AM (Mode C), while from left to right, we increase the value of $f$ in each case. As we see from the figure, with decreasing specific AM of the evaporative wind, the binary eventually evolves to a wider orbit, the duration of RLOF is reduced, and the companion star is more likely to be detached from its RL. Obviously increasing the value of $f$ results in a similar tendency. \citet{Smedley2015} showed that when $f\gtrsim 0.12$, evaporation is strong enough to keep the companion star detaches immediately in their AIC model. Our calculations indicate that whether the companion star can fill its RL heavily depends on the choice of the evaporation model: the donor star can detach immediately in some sequences of Modes B and C, while it can hardly detach after the start of evaporation in Mode A, even with a large value of $f$.
The evaporation efficiency also affects the formation of ultra-compact systems below 0.1 day. For contracting systems, after the cessation of MB, GR-induced AML acts as a weak counterweight to the orbital expansion, so ultra-compact systems can be formed only with weak evaporation. \cite{Smedley2015} showed that, when $f<0.044$ evaporation is not strong enough and the systems can reconnect for another phase of RLOF. Figure~5 reveals that this value of $f$ also depends on the adopted AML modes, and the orbital evolution is much more sensitive to the values of $f$ in Modes B and C than in Mode A.

\noindent (4) Will redbacks evolve to black widows?


According to \cite{Chen2013}, black widows and redbacks are two
distinct populations of eclipsing MSPs for different evaporation
efficiencies due to the geometric effect. On the contrary,
\cite{Benvenuto2014,Benvenuto2015} argued that redbacks with compact
orbits would evolve to black widows, while the ones with longer
orbital periods would evolve to MSP-He WD systems. Basically,
whether or not a redback can evolve to a black widow depends on two
factors: (1) the evaporation efficiency, and (2) the balance between orbital expansion caused by mass transfer/loss and orbital contraction induced by AM carried away by the evaporative wind.
Our calculations indicate that, for suitable choice of the $f$ values, (at least some of) compact redbacks can evolve to black widows while the wide ones will form MSP-He WD systems in both Modes A and B.

\section{SUMMARY}
This work is motivated by the recent discovery that the three
transitional pulsars are both redbacks and transient AMXPs. We then
suggest the thermal and viscous instability in the accretion disks
in LMXBs to be a possible mechanism for the formation of redbacks.
We investigate the parameter space for the occurrence of the disk
instability in the evolution of LMXBs, and the influence of
evaporation on their secular evolution. Our results can be
summarized as follows.

1. Almost all LMXBs are subject to the thermal and viscous
instability and experience transient phase(s) sooner or later in
their evolution. If the NSs have be spun up to millisecond periods,
the reduction of the mass transfer rate during quiescence provides a
plausible mechanism for switching on the pulsar activity and the
formation of redbacks, besides the disrupted MB and irradiation
instability models.

2. In this scenario, the redback companions can be MS stars, evolved
stars, and (proto-)WDs with either filled or underfilled Roche-lobes.

3. Considering the influence of evaporation on the binary evolution, some fraction of redbacks evolving along the contracting tracks may become black
widows, while the wide ones may form MSP-He WD binaries. The final
products are heavily dependent on the evaporation efficiency and the
specific AM carried away with evaporative mass loss, both of which deserve further study.

\begin{acknowledgements}
We are grateful to an anonymous referee for constructive comments, and Hailiang Chen and Thomas Tauris for helpful discussions. This work was supported by the Natural Science Foundation of China under grant Nos. 11133001 and 11333004, the Strategic Priority
Research Program of CAS under grant No. XDB09010200, and the Munich Institute for Astro- and Particle Physics (MIAPP) of the DFG cluster of excellence ``Origin and Structure of the Universe".

\end{acknowledgements}



\newpage
\begin{landscape}
\begin{figure}[h,t]
\plotone{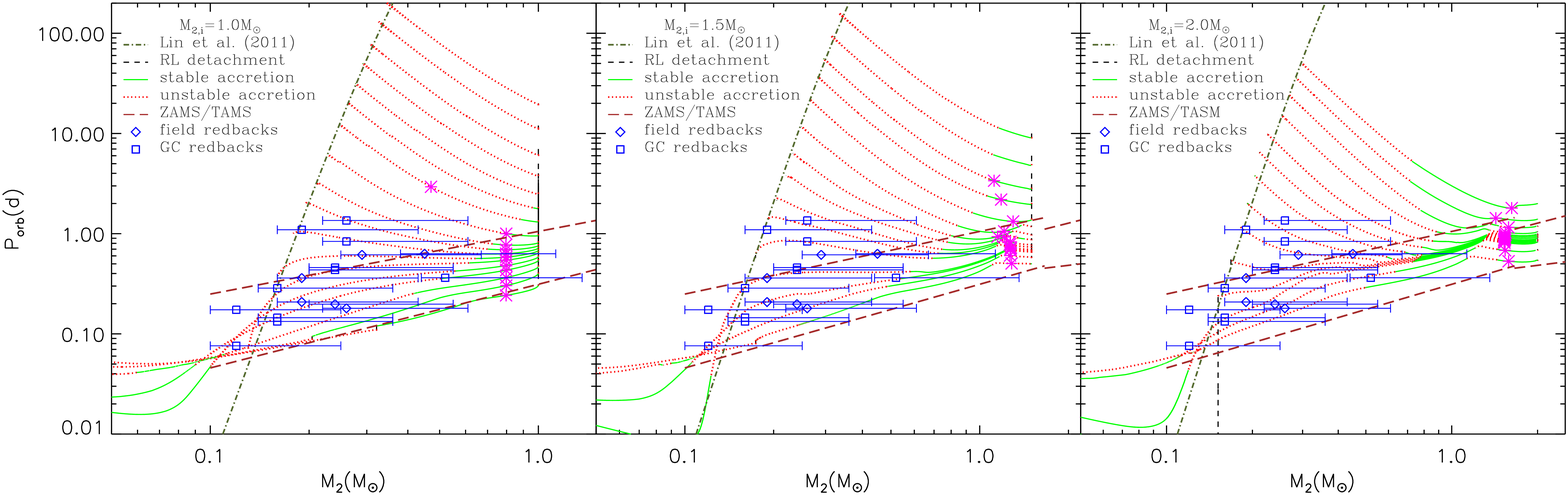} \caption{Evolutionary tracks of LMXBs in the
$M_{2}-P_{\rm orb}$ plane. From left to right the initial masses of
the companion stars are taken to be $M_{2, \rm i} = 1.0, 1.5, \rm
and~2.0~M_{\odot}$, respectively. The green solid, red dotted, and
black dotted lines represent the mass transfer processes with a
stable and unstable accretion disk, and the stage of RL detachment,
respectively. The dotted-dashed line shows the theoretical $P_{\rm
orb}-M_{\rm WD}$ relation for case B mass transfer. The asterisks on the evolutionary tracks
indicates the time when the NS has accreted $0.1M_{\odot}$ material.
The two \textbf{brown} dashed lines give the orbital periods with a ZAMS and
TAMS donor. Also plotted are redbacks in the Galactic disk and
globular clusters with blue symbols. \label{figure1}}
\end{figure}
\end{landscape}

\begin{figure}
\centering

{\label{main:a}\includegraphics[width=\textwidth,height=\textheight,keepaspectratio]{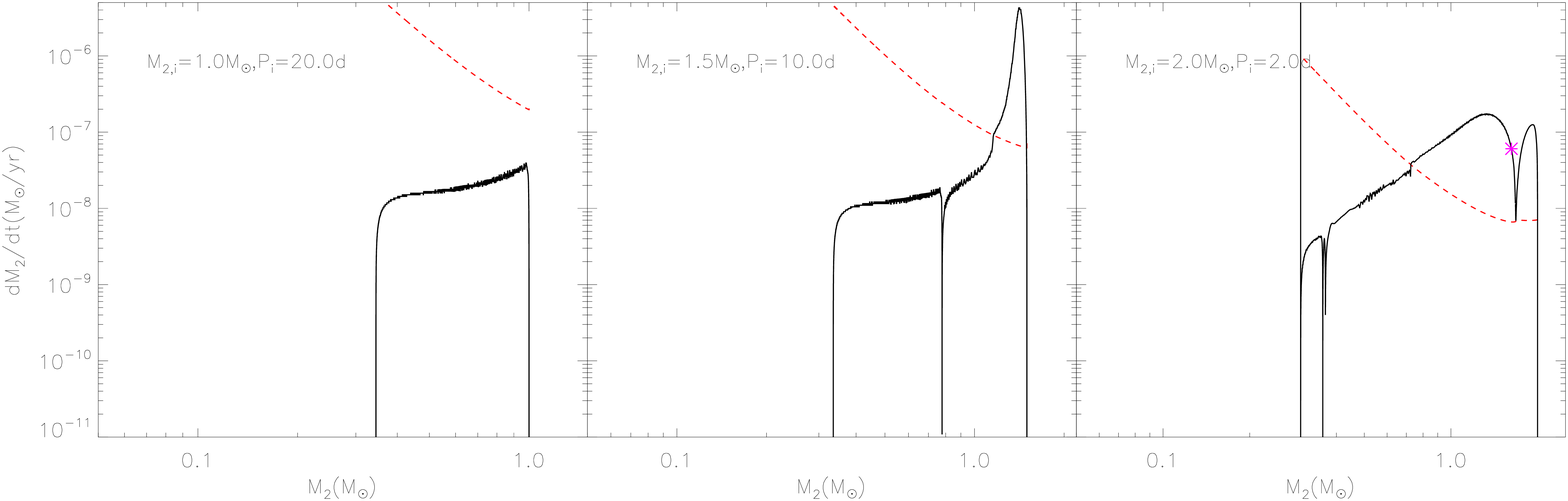}}
\centering

{\label{main:b}\includegraphics[width=\textwidth,height=\textheight,keepaspectratio]{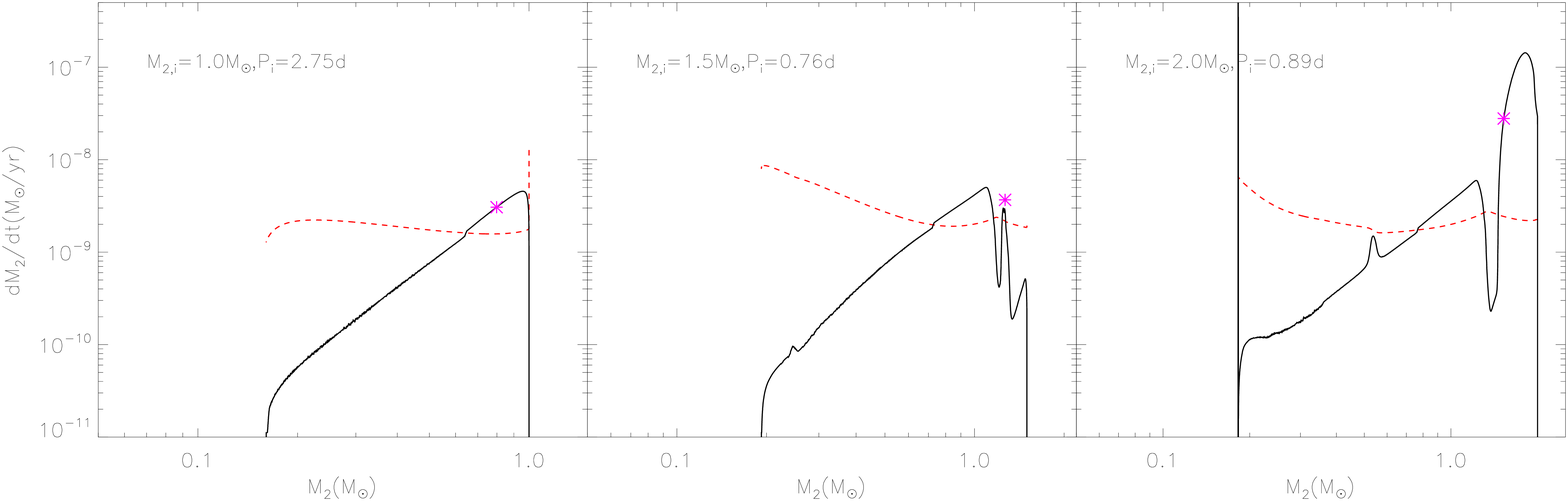}}
\centering

{\label{main:c}\includegraphics[width=\textwidth,height=\textheight,keepaspectratio]{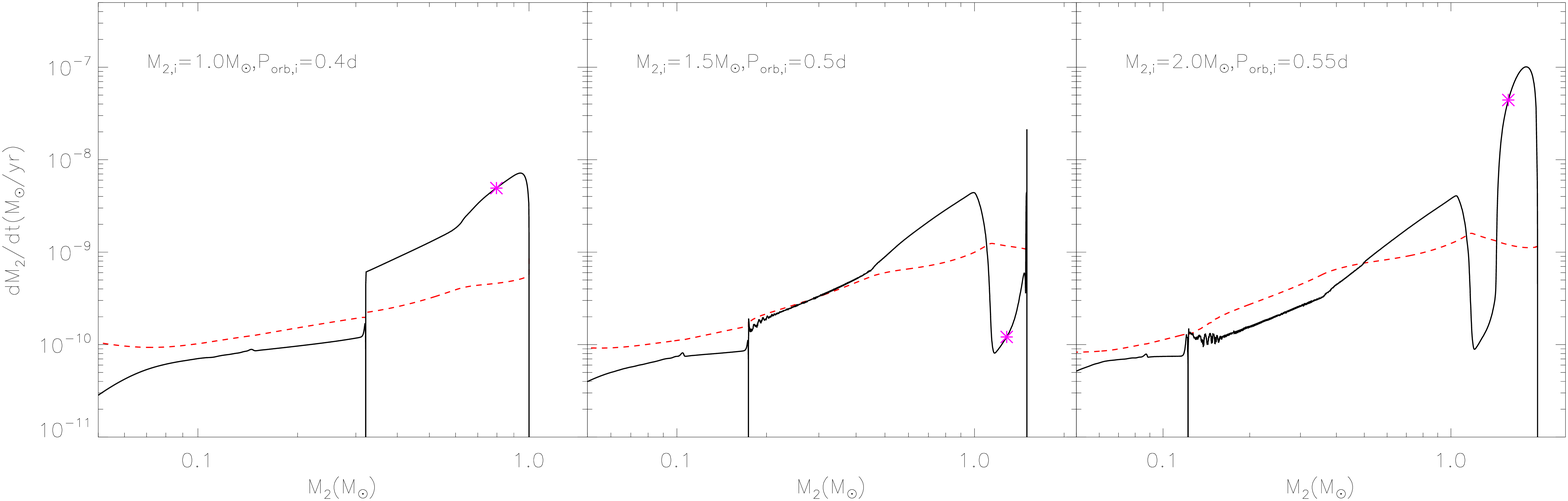}}
\caption{The mass transfer rate through RLOF (black solid line) and
the critical accretion rate for disk instability (red dashed line)
as a function of the donor mass. The bottom to top rows show
\textbf{contracting} to \textbf{widening} evolutionary tracks with
increasing initial orbital periods. In the left, middle, and right columns the initial donor masses are $M_{2, \rm i} = 1.0, 1.5, \rm
and~2.0~M_{\odot}$, respectively. The asterisk indicates when the NS has accreted $0.1~M_{\odot}$ material.} \label{figure2}
\end{figure}

\begin{landscape}
\begin{figure}[h,t]
\centerline{\includegraphics[width=1.3\textwidth,height=\textheight,keepaspectratio]{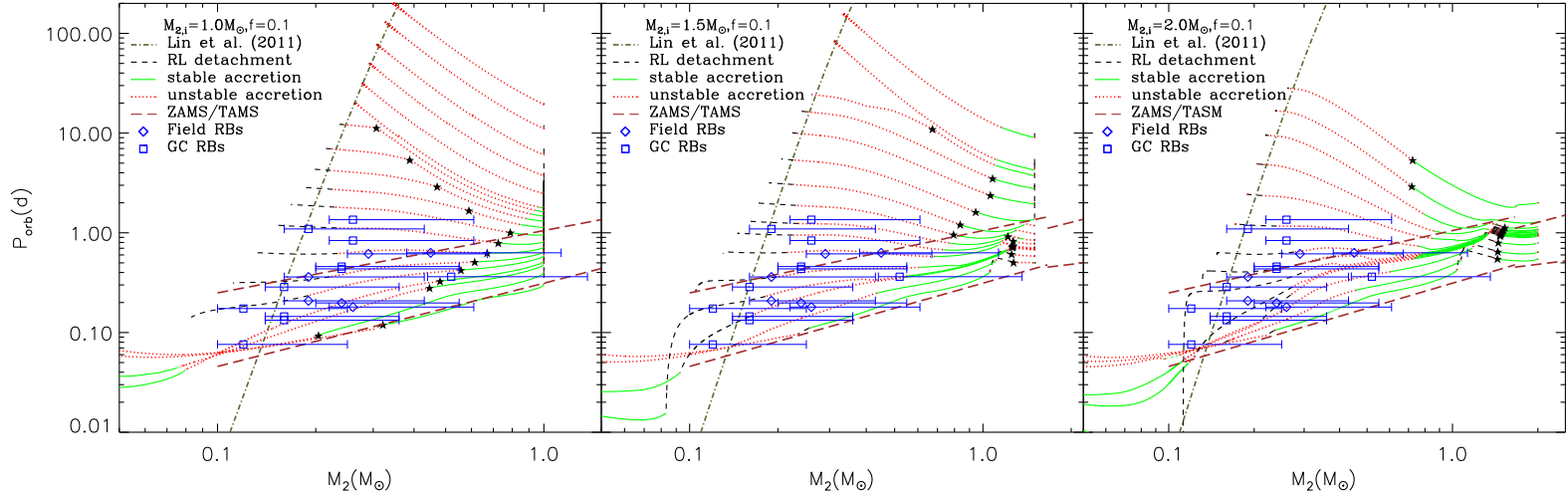}}
\caption{Similar to Fig.~1 but with evaporation taken into account
($f=0.1$). The beginning of evaporation is labeled with a solid
pentagram. \label{figure3}}
\end{figure}
\end{landscape}

\begin{figure}
\centering
{\label{main:a}\includegraphics[width=\textwidth,height=\textheight,keepaspectratio]{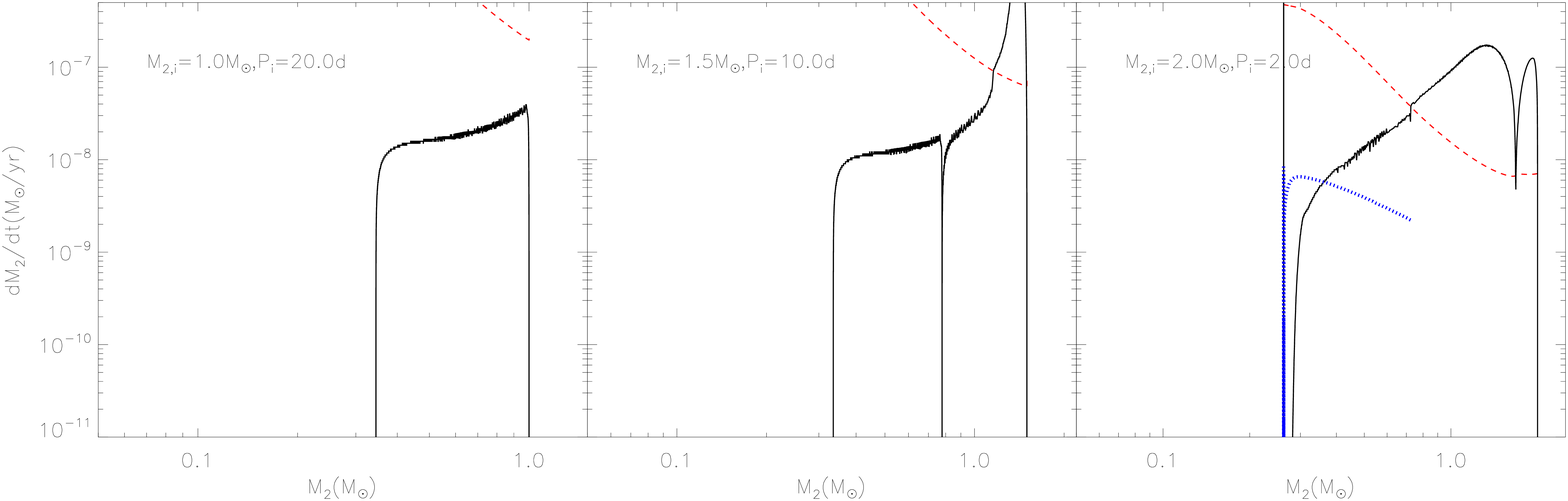}}
\centering
{\label{main:b}\includegraphics[width=\textwidth,height=\textheight,keepaspectratio]{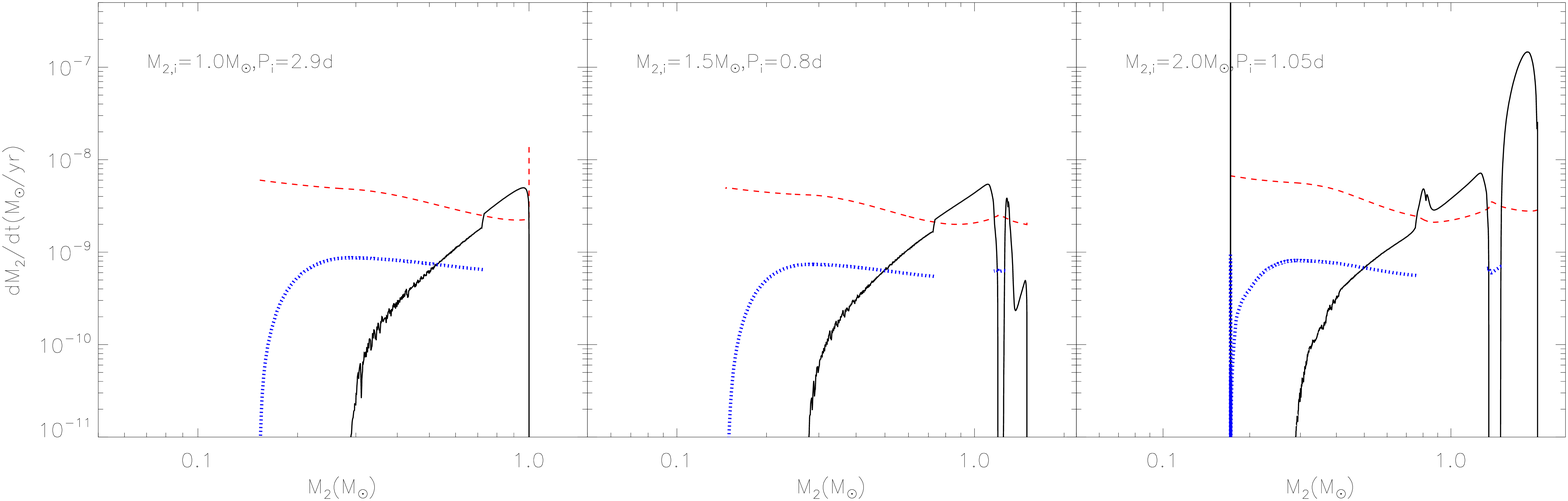}}
\centering
{\label{main:c}\includegraphics[width=\textwidth,height=\textheight,keepaspectratio]{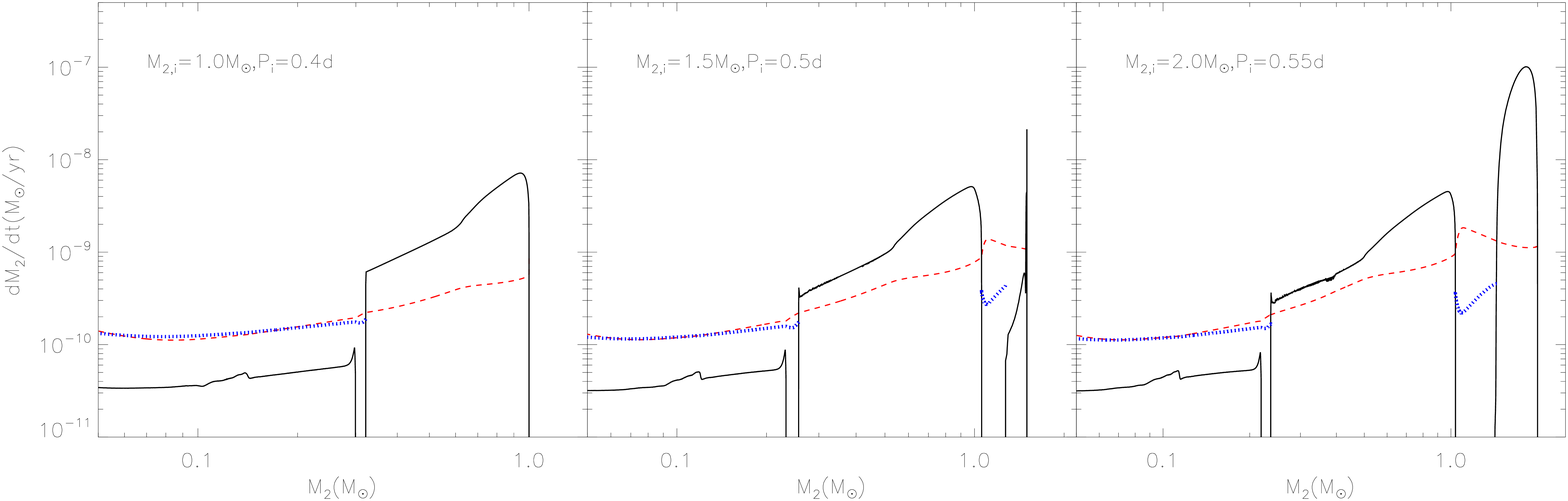}}
\caption{Similar to Fig.~2 but with evaporation taken into account
($f=0.1$). The mass loss rate caused by evaporation is plotted with
the blue dotted lines. \label{figure4}}
\end{figure}

\newpage
\begin{figure}
\centering
\label{main:a}
\plotone{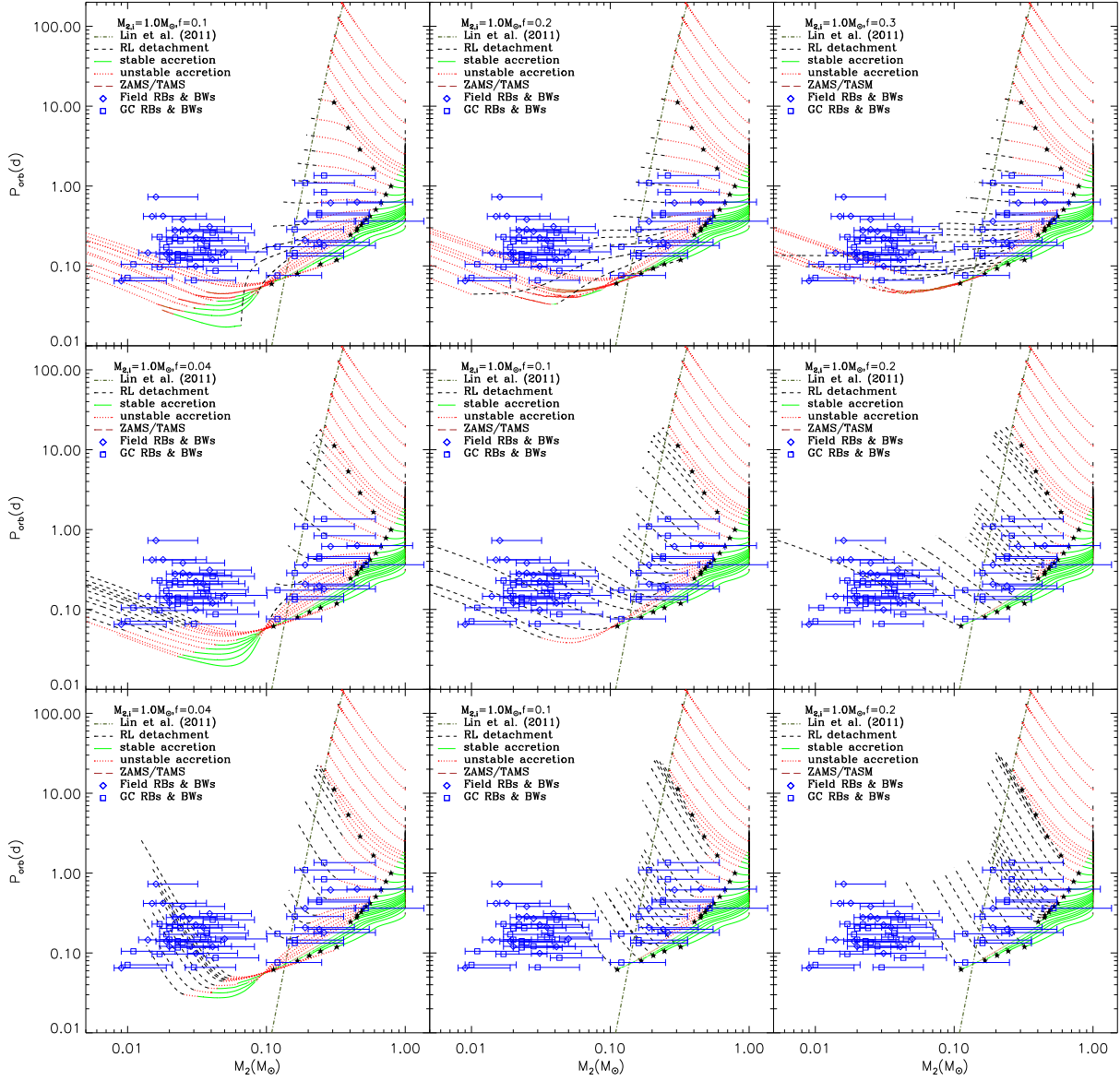}
\caption{Similar to Fig.~3 but with various efficiency and AML modes of the evaporative wind. From top to bottom rows, the evaporative wind is assumed to carry the specific angular momentum of the donor star (Mode A), at the inner Lagrangian point (Mode B), and no AM (Mode C) respectively. The evaporation efficiency is taken to be $f=0.1$, 0.2, and 0.3 for Mode A, and 0.04, 0.1, and 0.2 for Modes B and C.
Also plotted are black widow systems in the Galactic field (Field BWs) and
globular clusters (GC BWs) with blue symbols. \label{figure5}}
\end{figure}


\end{document}